# Update on the high speed serializer ASIC development for ATLAS Liquid Argon calorimeter upgrade

**Tiankuan Liu**[a], on behalf of the ATLAS Liquid Argon Calorimeter Group

[a] *Southern Methodist University,*
*Dallas, Texas 75275, USA*
*E-mail*: tliu@mail.smu.edu

ABSTRACT: We have been developing a serializer application-specific integrated circuit (ASIC) based on a commercial 0.25-μm silicon-on-sapphire (SOS) CMOS technology for the ATLAS liquid argon calorimeter front-end electronics upgrade. The first prototype, a 5 Gbps 16:1 serializer has been designed, fabricated, and tested in lab environment and in a 200 MeV proton beam. The test results indicate that the first prototype meets the design goals. The second prototype, a double-lane, 8 Gbps per lane serializer is under development. The post-layout simulation indicates that 8 Gbps is achievable. In this paper we present the design and the test results of the first prototype and the design and status of the second prototype.

# Contents

## 1. Introduction

The optical data links for the ATLAS liquid argon (LAr) calorimeter between the front-end boards (FEBs) and the back-end electronics operate at 1.6 gigabit per second (Gbps) per fiber channel [1-2]. In the LAr calorimeter readout electronics upgrade, it is proposed to remove the Level-1 trigger from FEB and transmit continuously digitized data off the detector. Consequently, the data rate of the optical links increases from 1.6 Gbps to about 100 Gbps per FEB [3-4]. However, the data rate of G-Link [5] or GOL [6] currently used in high energy physics experiments is not more than 1.6 Gbps, too slow for the upgrade. Thus for the LAr calorimeter readout electronics upgrade, a high speed radiation tolerant serializer ASIC is required.

     To meet this challenge, we have been developing a serializer application-specific integrated circuit based on a commercial 0.25-μm silicon-on-sapphire (SOS) CMOS technology. The first prototype, a 5 Gbps 16:1 serializer dubbed as LOCs1 [7-8] has been designed, fabricated, and tested in lab environment and in 200 MeV proton beam. The second prototype, a double-lane, 8 Gbps per lane serializer (dubbed as LOCs2) is under development. In this paper we present the design and the test results of LOCs1 and the design and status of LOCs2.

## 2. The design and test of a single-lane serializer

### 2.1 The design

A 5-Gbps 16:1 serializer, named as LOCs1 [7-8], based on a commercial 0.25-μm SOS CMOS technology has been prototyped. LOCs1 is the first step towards optical links for the readout upgrade of the ATLAS LAr calorimeter for the super LHC. The serializer consists of a serializing unit, a phase lock loop (PLL) clock generator and a current-mode-logic (CML) driver as shown in Figure 1(a). The serializing unit multiplexes 16 bit parallel low-voltage differential signaling (LVDS) data into a serial bit stream. The serializer unit extends 2:1 multiplexers to a 16:1 one with binary tree architecture. Only the last 2:1 multiplexer needs to be optimized to work at the highest speed or 2.5 GHz. Two complimentary 2.5 GHz clock signals are required to speed up the D-flip-flop in the last 2:1 multiplexer. To achieve good immunity from the single-event effects, we use large size transistors and static D-flip-flops in the whole design. With a 312.5 MHz reference clock input, the PLL clock generator provides 2.5 GHz, 1.25 GHz, 625 MHz, and 312.5 MHz clock signals to the serializing unit. A multiple-pass loop differential ring oscillator [7, 9-10] is used to boost the operating frequency of voltage control oscillator (VCO). The PLL loop bandwidth is programmable for adapting different reference clock qualities. The PLL can be configured to lock to either the rising or falling edge of the reference clock. This edge-selection feature is useful for the users to latch data with optimal timing. The CML driver can drive 50-Ω transmission lines.

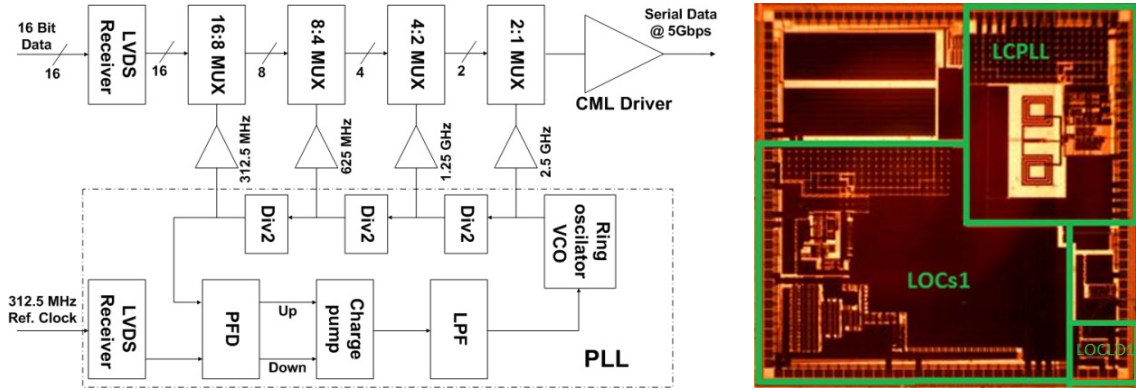

Figure 1: (a) Block diagram of the serializer LOCs1; (b) Die micrograph

The micrograph of the serializer ASIC is shown in Figure 1(b). The serializer occupies about 50% area of a 3×3 mm$^2$ die. All the I/O pins have electrostatic discharge protection except the high speed serial data output pins. The die also includes an LC-tank-based PLL (LCPLL) operating at near 5 GHz and a line driver/laser driver (LOCLD1). The LCPLL will be used in the next prototype of the serializer array. The LOCLD1 is used to drive 50-Ω coaxial cables and a laser diode.

### 2.2 The lab test

The test setup is shown in Figure 2(a). In the laboratory test, a field-programmable gate array (FPGA) based board provides 16 bit parallel data and a clock signal to a dedicated chip carrier board through a twisted pair cable. We measured the high speed serial data parameters through the SMA connectors on board with a high speed real time oscilloscope or a bit error rate tester. An eye diagram at 5 Gbps is shown in Figure 2(b).

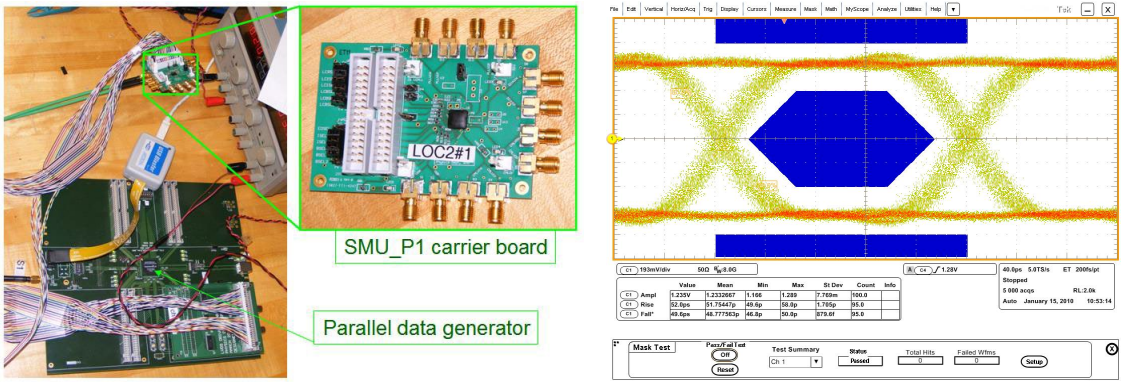

Figure 2: (a) Pictures of the test setup; (b) Eye diagram at 5 Gbps

The bit error rates (BER) of all the seven boards are better than $10^{-12}$ in the data range from 4.0 to 5.7 Gbps. The average values of the major measured parameters of the seven working boards are listed in Table 1. All parameters are measured at 5 Gbps except the upper and lower working data rate limits.

Table 1: Average values of the measured parameters of seven working boards

| Parameters | Measured results |
| --- | --- |
| Output Amplitude (peak-peak, V) | 1.16 |
| Rise time (20%−80%, ps) | 52.0 |
| Fall time (20%−80%, ps) | 51.9 |
| Total jitter at BER of $10^{-12}$ | 61.6 |
| Random jitter (RMS, ps) | 2.6 |
| Deterministic jitter (peak-peak, ps) | 33.4 |
| Eye opening at BER of $10^{-12}$ (ps) | 122 |
| Power consumption (mW) | 463 |
| Lower working limit (Gbps) | 4.0 |
| Upper working limit (Gbps) | 5.7 |

### 2.3 The radiation test

We have performed a radiation test with a 200-MeV proton beam at Indiana University Cyclotron Facility. The test setup is shown in Figure 3(a). A custom-made BER test system for online error detection was placed in an area shielded by lead bricks. We put two LOCs1 carrier boards in the beam and another one in the shielded area as a reference. To test the possible angle effects, the angle between the beam incident direction and die surface normal was set at 0, 30, 45, or 60 degree during the radiation test. The boards accumulated 90% of the total fluence when their angles were kept at 60 degree. Because the number of the single-event upsets (SEUs) were small, we did not observe any statistically significant dependence on angles.

The radiation test lasted for 12 hours in the beam and we kept the test system running for 15 hours after the beam off. We did not observe any bit error in the annealing time. We monitored the power supply current of the serializers during the test. The current change is shown in Figure 3(b). The currents changed less than 6% during the beam time and annealing time. This means the total ironizing dose (TID) effects are negligible for our application. We observed two types of SEUs: single bit errors and synchronization errors. We observed five single bit error events in total which did not affect the link status afterwards. The extrapolated BER for the single bit errors is $1.6\times10^{-18}$ at sLHC ATLAS LAr calorimeter. When a synchronization error event occurred, there were a burst of bit errors in a short duration. After the burst of bit errors, the received data had one bit shift comparing to the generated PRBS data for error checking in the error detector. The bit shift was removed when the receiver was reset for a word alignment. This burst of bit errors lasts only several tens bits, for each synchronization error event, the link can be recovered on the receiver side without many bit loss. The extrapolated number of synchronization error events is less than three at the ATLAS LAr calorimeter in the whole sLHC lifetime.

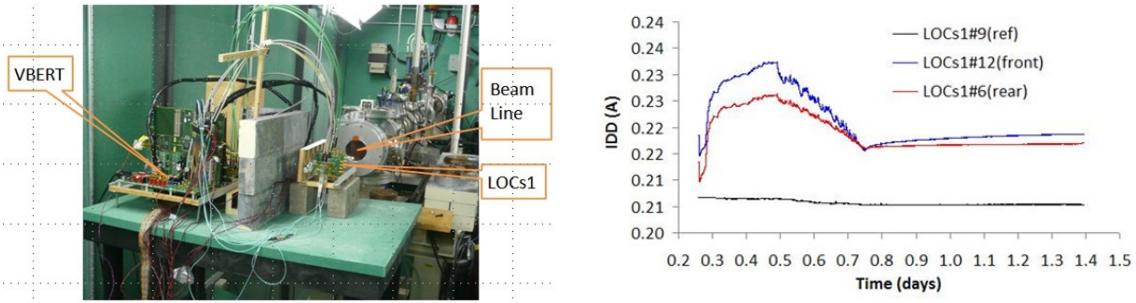

Figure 3: (a) A picture of the setup of radiation test; (b) The power supply current change during the test and during annealing time

## 2.4 The LCPLL

A low power, low jitter LCPLL [11-12] is also implemented in the same prototype as LOCs1. The LCPLL is the prototype of the high speed and low jitter clock generator for the next generation of serializer ASIC. The block diagram of the PLL is shown in Figure 4(a). An LVDS receiver and a CML driver are added as the input and output interface for test purpose. In the block diagram, the PFD is a phase and frequency detector. The charge pump (CP) converts the up and down signals into control current. The low pass filter (LPF) integrates the current into control voltage. The VCO is a LC-tank-based VCO. The divider chain consists of four divide-by-2 dividers, a high speed CML divider and 3 CMOS dividers. Because the bandwidth of the CML driver is not high enough, we monitor the output of the CML divider rather than the output of VCO. The output magnitude of a CML divider is not large enough to drive a CMOS divider and a CML driver, so a CML to CMOS converter is used after the CML divider.

The LCPLL has been characterized in laboratory environment. Figure 4(b) is the waveforms in which the output locked its phase to the input clock. Random jitter and deterministic jitter are about 1.3 ps and 7.5 ps, respectively. The measured tuning range, from 4.6 to 5.0 GHz, is narrower than the expected one which is from 3.8 to 5.0 GHz. The narrow tuning range issue has been investigated and understood. The power consumption at the central frequency is 111 mW at 4.9 GHz, comparing to 173 mW at 2.5 GHz of the ring oscillator based PLL used in the 16:1 serializer.

Two LCPLLs have been tested in a 200 MeV proton beam. The parameters of the irradiated PLLs were measured and compared with those of fresh PLLs. The output amplitude and power dissipation of these PLLs after irradiation increases about 25% and 9% of the average of the PLLs without irradiation, respectively, but the statistic is still very low to draw any conclusions. We will investigate this issue, although as a digital circuit the increase is not significant. There is no significant change in transition times and jitter performances. Both irradiated PLLs are functioning after the test indicating that these PLLs survived the TID test.

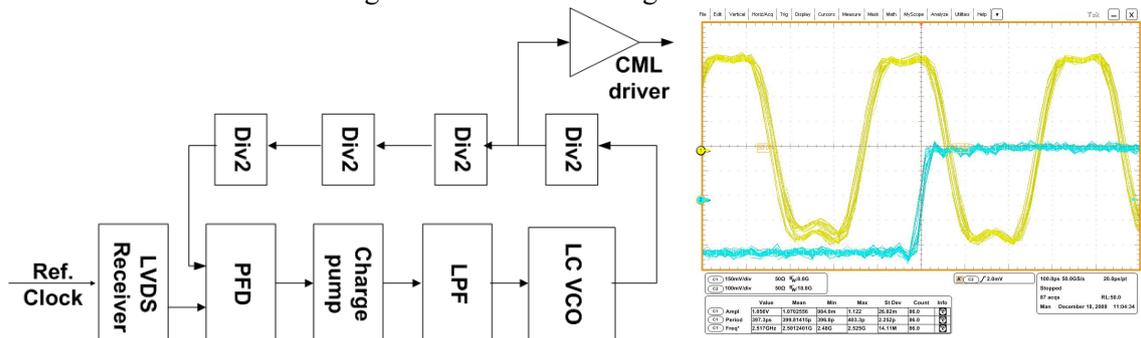

Figure 4: (a) block diagram of the LCPLL; (b) Waveforms of the PLL output clock (blue) locked to the input clock (yellow)

## 3. The design of a double-lane serializer and a laser driver array

The second prototype, dubbed as LOCs2, of the serializer ASIC is under development. The second prototype will include double serializer lanes, a shared LCPLL, and a VCSEL driver array. The block diagram of LOCs2 is shown in Figure 5. Each serializer lane operates at 8 Gbps and has 16-bit LVDS parallel input data and 1-bit CML serial output data. The two lanes share one LVDS clock input and an LCPLL. The basic architecture and low speed CMOS circuits used in the serializers and LCPLL are inherited from the first prototype. The VCSEL driver array will take the output signals of four serializer lanes and drive four VCSEL's at 8 Gbps.

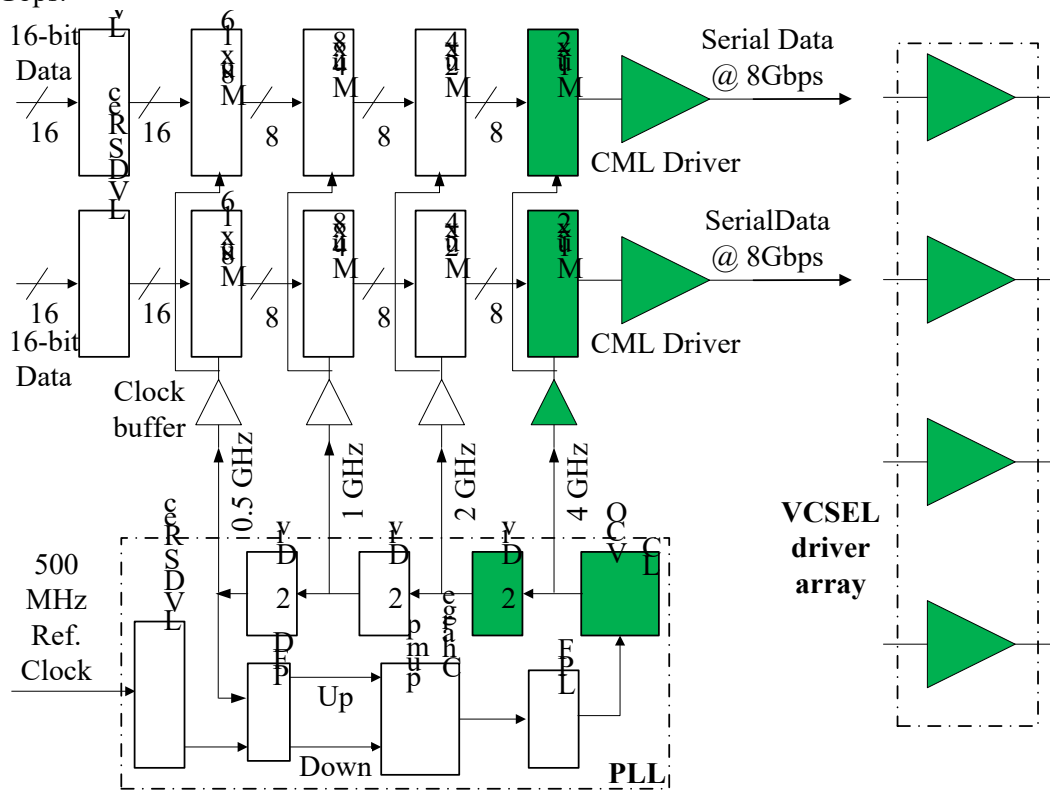

Figure 5: The block diagram of LOCs2

**3.1 The design of the double-lane serializer**

The designs of all fast parts in CML logic (green shapes in Figure 5) have been finished. The LCPLL implemented and tested in the first prototype will be used in LOCs2 with small tuning range modification to cover the frequency of 4 GHz. A bug of the divide-by-2 divider used in the first prototype LCPLL has been identified and a new divider has been designed. The post simulation shows that the maximum operation frequency is larger than 4.2 GHz in the worst case (the slow-slow corner and 85 °C). The clock buffer is used to fan out the high speed clock. The post layout simulation shows that the maximum operation frequency is larger than 5.4 GHz in the worst case (the slow-slow corner, 85 °C). The last stage of 2:1 multiplexer has been changed from the CMOS logic in the first prototype to the CML logic in the second prototype. Deterministic jitter is simulated in post layout to be about 4.5 ps with 8 Gbps PRBS (27-1) signals (typical corner and 27 °C). The plain resistor-load CML driver used in first prototype has been replaced with an active shunt peaking CML driver. Deterministic jitter is simulated in post layout is about 5 ps with 8 Gbps PRBS (27-1) signals in the typical corner and 27 °C.

**3.2 The design of a laser driver array**

The four-lane vertical-cavity surface-emitting laser (VCSEL) driver array is under development and will be prototyped in the same time as the double-lane serializers. We have finished the design of a single VCSEL laser except the biasing block.

The minimum input amplitude of the laser driver is set to be 200 mV (differential), the minimum output amplitude of the serializer. The output amplitude of the VCSEL driver is set by the laser diode. We set the output amplitude to be 600 ~ 800 mV over 100-Ω load, corresponding to the modulation current of 6 ~ 8 mA over a VCSEL through a pair of 50-Ω (differential 100-Ω) transmission lines. The biasing current will be set at 5 ~ 10 mA.

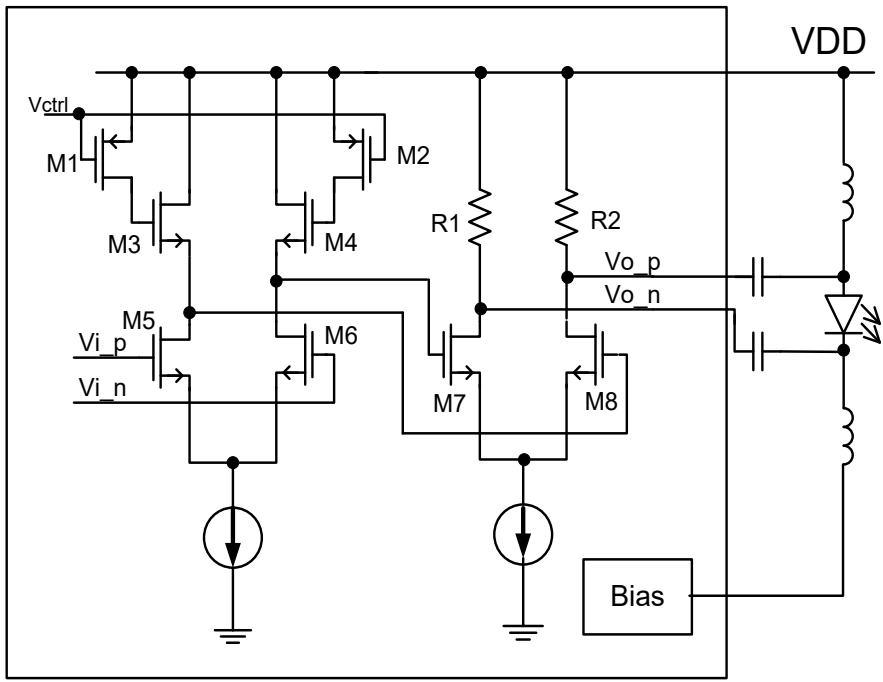

Figure 6: the schematic of the laser driver (only the last two stages shown)

The schematic of the laser driver is shown in Figure 6. Only the last two stages are shown in the figure. Due to the bandwidth limit of the process we use, a single stage cannot achieve the high output amplitude at the required data rate. The structure of the last stage depends on how the laser driver is connected the VCSEL. We use internal 50-Ω pull-up resistors (R1 and R2) and the AC coupling. The switch nMOSFET's (M7 and M8) are very large (200.6 μm wide, 0.25 μm long), meaning large capacitive loads for the previous stage. To drive this large capacitive load at a high frequency, we have to use multiple stages. We call the output stage as the main drive stage, and the stages before the main drive stage as the pre-drive stages. In our design, we have six pre-drive stages and one main drive stage. All pre-drive stages have the same structures.

Even with multi-stage structure, we still have to use a bandwidth extension technique. We choose the shunt peaking technique [13-14]. Since we cannot afford a pair of embedded spiral inductors (250 μm × 250 μm in the process we use) for each stage, we choose the active shunt peaking technique. nMOSFET's M3 and M4 are used as pull up resistors. pMOSFT's M1 and M2 operate in the triode region and act as resistors. The equivalent impedance of M3 and M1 is a resistor in series with an inductor. It is true for M4 and M2, too. The equivalent inductance is used to implement the shunt peaking effect. We use an external adjustable voltage (Vctrl) to adjust the peaking strength.

The eye diagram of the CML driver and the laser driver in the post layout simulation at the typical process corner and 27 °C is shown in Figure 7. The deterministic jitter is less than 6 ps, including the influence of the CML driver.

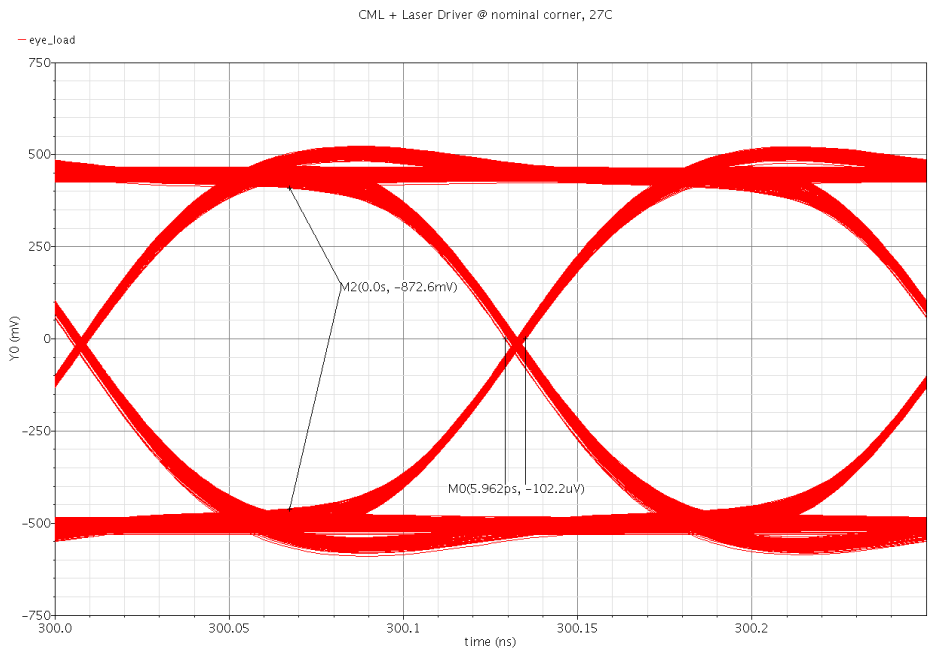

Figure 7: an eye diagram at 8 Gbps based on the post layout simulation

## 4. Conclusion

To meet the challenge of the ATLAS LAr calorimeter upgrade, we have been developing a serializer ASIC based on a commercial 0.25-μm SOS CMOS technology. The first prototype, a 5 Gbps 16:1 serializer has been designed, fabricated, and tested in lab environment and in 200 MeV proton beam. The test results indicate that the first prototype meets the design goals. The second prototype, a double-lane, 8 Gbps per lane serializer is under development. The post layout simulation indicates that 8 Gbps is achievable. The second prototype will be submitted in the beginning of 2012.

**Acknowledgments**

This R&D project is supported by US-ATLAS. We would like to thank Jasoslav Ban at Columbia University, Paulo Moreira at CERN, Fukun Tang at University of Chicago, Mauro Citterio and Valentino Liberali at INFN, Carla Vacchi at University of Pavia, Christine Hu and Quan Sun at CNRS/IN2P3/IPHC, Sachin Junnarkar at Brookhaven National Laboratory, Mitch Newcomer at University of Pennsylvania, Jay Clementson, Yi Kang, John Sung, and Gary Wu at Peregrine Semiconductor Corporation for their invaluable suggestions and comments to help us complete the design work. We also would like to thank Justin Ross at Southern Methodist University for helping us set up and maintain the design environment, Charles Joseph Nelson and Barbara von Przewoski at IUCF for their help in the radiation test.